\documentclass{iau}

\usepackage{amsmath}
\usepackage{graphicx}
\usepackage{multirow}

\begin{document}

\lefttitle{W. E. Banda-Barrag\'an, A.Antipov, and D. Villarruel}
\righttitle{Survival and synthetic observables of neutral atomic hydrogen in galactic wind simulations}

\jnlPage{1}{7}
\jnlDoiYr{2024}
\doival{10.1017/xxxxx}

\aopheadtitle{Proceedings IAU Symposium}
\editors{D.J. Pisano, Moses Mogotsi, Julia Healy, Sarah Blyth, eds.}

\title{Survival and synthetic observables of neutral atomic hydrogen in galactic wind simulations}

\author{Wladimir E. Banda-Barrag\'an$^{1,2}$, Andrei Antipov$^3$, and Daniel Villarruel$^{1}$}

\affiliation{$^1$Escuela de Ciencias F\'isicas y Nanotecnolog\'ia, Universidad Yachay Tech, Hacienda San Jos\'e S/N, 100119 Urcuqu\'i, Ecuador (E-mail: \href{mailto:wbanda@yachaytech.edu.ec}{wbanda@yachaytech.edu.ec})\\
$^{2}$Hamburger Sternwarte, University of Hamburg, Gojenbergsweg 112, 21029 Hamburg, Germany\\
$^3$Centre for Astrophysics and Planetary Science, Racah Institute of Physics, The Hebrew University, Jerusalem 91904, Israel}

\begin{abstract}
Connecting numerical simulations to observations is essential to understanding the physics of galactic winds. Our Galaxy hosts a large-scale, multi-phase nuclear wind, whose dense gas has been detected using H\,{\sc i} and molecular line observations. In this paper, we summarise our recent numerical work devoted to producing synthetic H\,{\sc i} observables and measuring the properties of H\,{\sc i} gas in galactic wind simulations. We discuss the evolution of radiative cloud systems embedded in star formation-driven galactic winds. Our shock-multicloud models show that multicloud gas streams are able to produce significant fractions of H\,{\sc i} gas via recondensation. Our wind-cloud models show that magnetic fields have significant effects on the morphology and spectral signatures of H\,{\sc i} gas. Cooling-driven recondensation, hydrodynamic shielding, and magnetic draping promote the survival of dense gas and the development of filamentary outflows. The orientation of magnetic fields also has an effect on synthetic observables, particularly on H\,{\sc i} spectral lines. Transverse magnetic fields produce broader spectral lines of H\,{\sc i} than aligned magnetic fields. Our models and analysis suggest that the fast-moving H\,{\sc i} gas observed in the nuclear wind of our Galaxy may arise from multi-phase flows via recondensation.
\end{abstract}


\maketitle

\section{Introduction}
Studying galactic winds is important to developing a comprehensive understanding of galaxy evolution (\citealt{2024arXiv240608561T}). Galactic winds are produced via feedback processes in galaxies, e.g. star formation (SF) or active galactic nuclei (AGN). They regulate the cosmic cycle of matter (\citealt{2017ARA&A..55..389T}) by promoting the exchange of metals between the interstellar (ISM) and circumgalactic medium (CGM) of the host galaxies and their external intergalactic medium (IGM). Our empirical understanding of these structures comes from emission and absorption line studies of the CGM (\citealt{2024arXiv240603553F}). Observational studies show that galactic winds are multi-phase media and have a complex multi-scale substructure (\citealt{2024AJ....168...11K}).\par

Our Galaxy, for instance, is known to host a large-scale, multi-phase nuclear wind (\citealt{2024A&ARv..32....1S}). This outflow has been detected at multiple wavelengths and at several scales (\citealt{2019Natur.567..347P}). At very large scales the so-called Fermi bubbles are visible in gamma-ray emission (\citealt{2010ApJ...724.1044S}), extending $\sim 8\,\rm kpc$ above and below the Galactic plane. A radio counterpart extending even farther away has also been found (\citealt{2013Natur.493...66C}). At smaller scales, chimneys extending 10's to 100's of pc have also been detected in X rays (\citealt{2021A&A...646A..66P}) and molecular line observations (\citealt{2023A&A...674L..15V}). At intermediate scales, ranging from $\sim 0.4$ to $\sim 1\,\rm kpc$ populations of high-velocity H\,{\sc i} gas (\citealt{2013ApJ...770L...4M}), as well as molecular gas (\citealt{2020Natur.584..364D}), have been found populating the base of the Fermi bubbles.\par

The relation between these outflow signatures, the connection between multiple scales, the wind-launching mechanisms (e.g. SF- versus AGN-driven feedback), and the overall structure of this GC outflow are all open questions. Answering these questions requires connecting the results of numerical simulations to observations. Thus, in this paper we summarise our recent efforts in developing tools to study the properties of dense gas (H\,{\sc i} gas, in particular) from numerical simulations.

\section{Galactic wind simulations}
To study the physics of H\,{\sc i} gas we resort to idealised numerical simulations, which we run with the PLUTO code (\citealt{2007ApJS..170..228M}). In this paper, we briefly discuss two types of simulations, which are designed to extract information on dense gas embedded in galactic winds, at different physical scales and with distinct levels of complexity. Our models are:\vspace{-0.2cm}

\begin{figure}[h!]\centering
  \includegraphics[scale=.36]{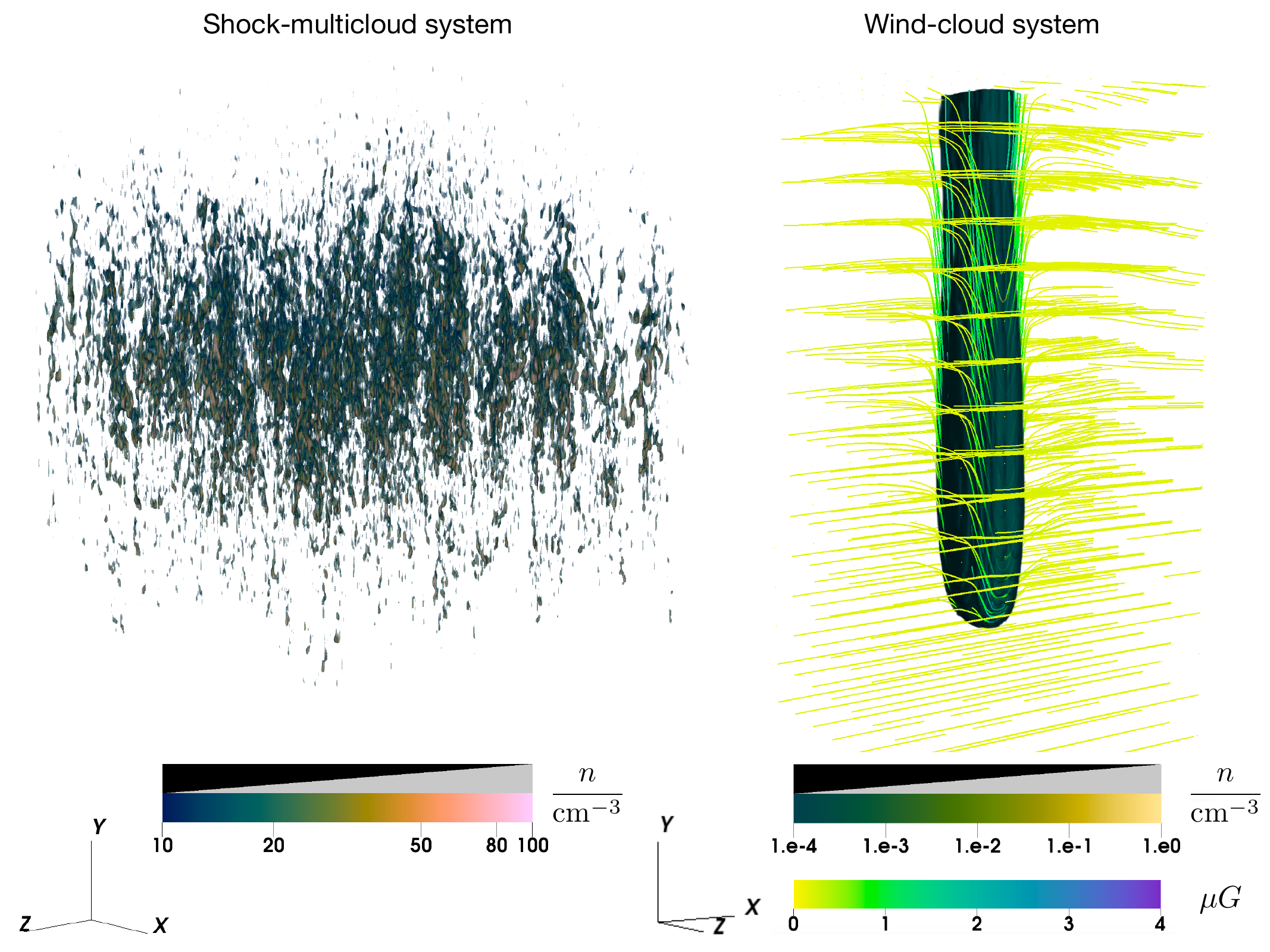}
  \caption{Left: Number density rendering from a shock-multicloud simulation showing the filamentary structure of dense gas in a multi-phase outflow (see \citealt{2021MNRAS.506.5658B,2025Antipov}). Right: snapshot of a wind-cloud model showing the cloud number density and the 3D topology of a transverse magnetic field that drapes around the cloud (see \citealt{2024A&A...689A.127C}).}
  \label{Fig1}
\end{figure}

\subsection{Shock-multicloud models}
Our shock-multicloud models (Figure \ref{Fig1}) are designed to study the interactions between a hot, fast-moving outflow and a layer of cold, initially static clouds (\citealt{2020MNRAS.499.2173B,2021MNRAS.506.5658B}). The initial conditions comprise a strong $M=10$ shock, a background medium, and a dense gas layer with a log-normal density distribution that follows a turbulent power-law spectrum. As the shock travels through the background medium, it leaves behind an over-pressurised post-shock flow (representing a wind), which interacts with the clouds in the layer. The dynamical instabilities that arise at the cloud-shocked gas interface due to velocity shear then disrupt and ablate dense gas, leading to its efficient mixing with the background and wind gas. This mixed gas is thermally unstable as it acquires temperatures at which radiative cooling is very efficient. As a result the gas cools down and forms a multi-phase medium. Some of the cooling gas sits at $\sim 10^4\,\rm K$, while some of the denser regions can also cool further down to temperatures near the cooling floor of our simulations at $\sim 10^2\,\rm K$. The end result is a three-phase medium where the warm and cold phases are spatially coincident and localised at filamentary cloudlets with low volume filling factors (see \citealt{2025Antipov}), and the hotter phases are more diffuse and have much larger volume filling factors. 

\subsection{Wind-cloud models} 
Our wind-cloud models aim at studying the micro physics of interactions between dense and diffuse gas in galactic winds (\citealt{2018MNRAS.473.3454B,2024A&A...689A.127C}). While these single cloud models are more idealised than multicloud models, they allow to resolve mixing layers with a higher resolution and study the effects of magnetic fields on clouds in much greater detail. In fact, wind-cloud models have revealed important aspects of the physics of dense gas. For instance, it has been extensively proven that adiabatic clouds can be readily destroyed by Kelvin-Helmholtz (KH) and Rayleigh-Taylor (RT) instabilities, but radiative cooling, thermal conduction and magnetic fields can have protective effects in certain conditions. When cooling is efficient, cloud gas can survive ablation for longer (\citealt{2009ApJ...703..330C}). Electron thermal conduction can create a transition layer around high-column density clouds (\citealt{2016ApJ...822...31B}). In the super-Alfvenic regime, magnetic fields transverse to the stream-wise motion can drape around clouds and shield dense gas (\citealt{2020ApJ...892...59C}). Our magnetohydrodynamical (MHD) wind-cloud models confirm these earlier findings, and show that transverse magnetic fields can envelope clouds compressing them in the direction of draping and stretching them in the direction perpendicular to it (see \citealt{2024A&A...689A.127C}).

\section{Survival and synthetic observables of H\,{\sc i} gas}
Using data from the previously-described simulations, we can study the properties of dense gas. In this section we describe how dense gas can survive in galactic winds, and what effects magnetic fields have on synthetic observables. Here we place particular attention to H\,{\sc i} gas owing to the topic of the IAUS 392 symposium, but more general results for other ions and gas phases can be found in our papers. The reader is referred to \cite{2025Antipov} (submitted) and \cite{2024MNRAS.535.1163V} for the properties of dense gas in multicloud models, and \cite{2024A&A...689A.127C} and our accompanying poster paper \cite{2024Villarruel} for the effects of magnetic fields and UV backgrounds, respectively, on clouds and synthetic spectral lines.


\begin{figure}[htbp]
    \centering
    \begin{tabular}{c c}
    \hspace{-0.35cm}\includegraphics[width=0.53\textwidth]{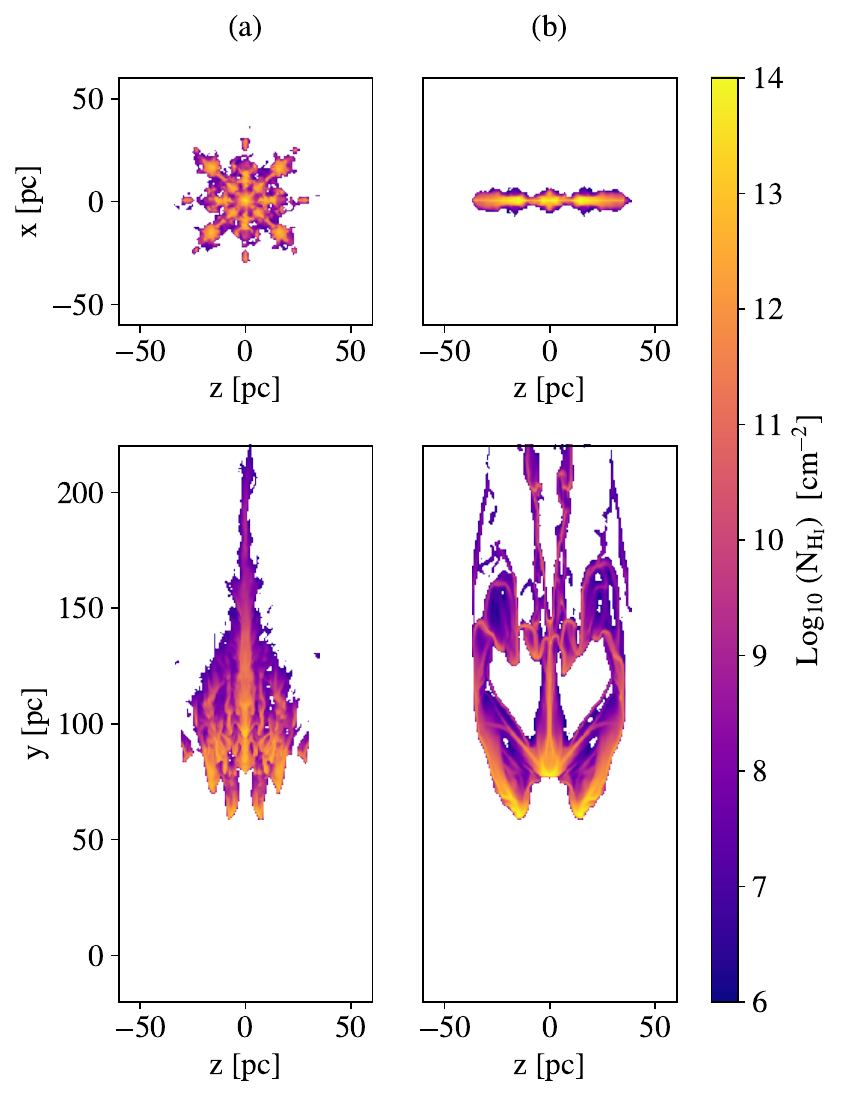} & 
    \hspace{-0.35cm}\includegraphics[width=0.47\textwidth]{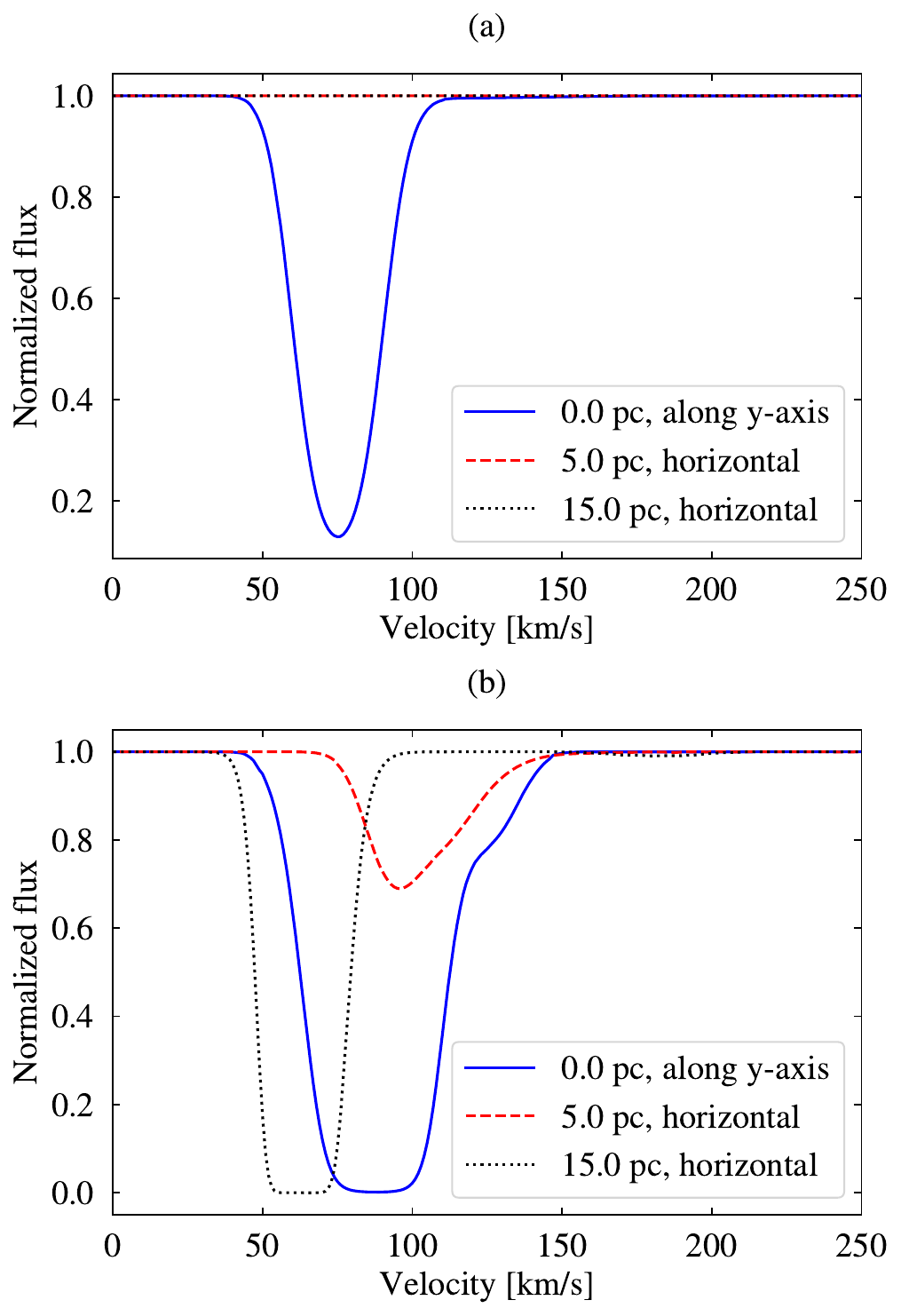}\\
    \end{tabular}
    \caption{Left: Synthetic column densities for H\,{\sc i} for both orientations of the magnetic field (a: aligned field, b: transverse field) at a distance of $50$ kpc from the starburst at $2.2$ Myr. Right: H\,{\sc i} absorption spectral lines for both models (a: aligned field, b: transverse field) obtained from our new python-based synthetic spectra generator (see our accompanying paper, \cite{2024Villarruel} for further details).}
    \label{Fig2}
\end{figure}

\subsection{Survival of dense gas in galactic winds} 
First, we discuss three mechanisms that our shock-multicloud and wind-cloud models have revealed as key players in preserving dense gas (including H\,{\sc i} gas) in multi-phase winds, namely: recondensation, hydrodynamic shielding, and magnetic draping.

\begin{enumerate}
    \item \textbf{Recondensation}: Our models show that multi-phase gas is the result of the interplay between radiative cooling, radiative heating, shock-driven heating, and turbulence. In these models pristine dense gas does not survive. It becomes mixed with the background flow, and then recondenses back. The new recondensed dense gas is still present at late stages, and we find that some of it comes from the hot wind, which implies mass growth. As a result, the cold gas detected in observations of e.g. the nuclear wind in our Galaxy, including H\,{\sc i}-emitting, H\,{\sc i}-absorbing, and molecular gas, is very likely gas that has condensed back from fast-moving, mixed, warm gas. Recondensation then implies that dense gas can be recycled along outflows and can acquire high velocities, not just from the wind ram pressure directly, but from the mixed phase that forms following the formation of mixing layers. H\,{\sc i} gas is produced self-consistently in multi-phase outflows. The cloudlets that form are enveloped by warmer filamentary shells (\citealt{2025Antipov}). Unresolved, cold ‘molecular’ cores spatially coincide with the location of H\,{\sc i} cloudlets, which is promising. In addition, hot gas has high volume filling factor $>0.9$, but warm $\sim10^4\,\rm K$ and cold $\sim 10^2\,\rm K$ gas has high mass content $\sim0.7$.
    \item \textbf{Hydrodynamic shielding}: Another interesting effect of multicloud systems is the protective effects of cloud streams. Hydrodynamic shielding refers to the capability of clouds moving along a stream to protect and shield themselves from hydrodynamic drag. Our controlled numerical experiments show that hydrodynamic shielding operates in supersonic flows when the cooling length is shorter than the cloud radius (\citealt{2024MNRAS.535.1163V}). Drag forces are reduced when we have systems of clouds with small separation distances. This also reduces the size of the mixing cylinder around each cloud and as a consequence the disruptive effects of shear instabilities on them. While our models are idealised as we include spherical clouds placed along a vertical stream, they illustrate well this effect, which is expected in more complex configurations and in realistic galactic winds. Indeed, when winds are launched from the ISM, the bubbles and shocks initially interact with cloud complexes in close configurations. The observational significance of our results is that the survival of dense phase in outflows (H\,{\sc i} inclusive) is very likely aided by hydrodynamical shielding too (in addition to recondensation).
    \item \textbf{Magnetic draping}: Our wind-cloud models (\citealt{2024A&A...689A.127C}) confirm earlier findings on the protective role of transverse magnetic fields. Even in cases where magnetic fields are weak (i.e. in super-Alfvenic cases), the field lines in transverse cases can envelope dense gas, thus creating a thin magnetic layer around the cloud (\citealt{2020ApJ...892...59C}). The net effect of magnetic draping is to shield the clouds from the wind by reducing dynamical instabilities, and to create a bumper at the leading edge of cloud (\citealt{2000ApJ...543..775G}), which can promote the growth of RT instabilities. Owing to the resulting (asymmetric) magnetic stresses the cloud is deformed and acquires a sheet-like morphology. The observational impact of magnetic draping is discussed in the section below.
\end{enumerate}

\subsection{H\,{\sc i} spectral lines from wind-cloud models} 
While the aforementioned effects of magnetic fields are well known in the literature, our models and analysis extend previous work (including our own) to study the spectral signatures of magnetic fields on H\,{\sc i} gas (see \citealt{2024Villarruel}). We investigate the spectral changes associated with the morphological effects that magnetic fields have on two cloud models: one with an aligned field and one with a transverse field. Figure \ref{Fig2} shows the influence of magnetic fields on the column densities and absorption spectra of H\,{\sc i} gas. Our analysis shows that, in addition to morphological changes, magnetic draping can also leave observational imprints in the form of broader spectral lines in the transverse field case, compared to the aligned case.  Overall, these results imply that H\,{\sc i} spectral lines in observations can also carry information on the underlying magnetic properties of outflowing clouds.

\section{Conclusions}
In this paper, we have briefly summarised our recent numerical work aimed at studying the properties and synthetic observables of H\,{\sc i} gas in galactic wind simulations. We have discussed two types of simulations, namely shock-multicloud and wind-cloud models, which reveal important insights into the physics of the population of H\,{\sc i} clouds in the nuclear wind of our Galaxy. Our conclusions are the following: 

\begin{itemize}
    \item Our shock-multicloud models show that multicloud gas streams are able to survive and produce significant fractions of H\,{\sc i} gas. We find that while dense gas is destroyed and mixed in these winds, this very same mixed gas is the seed for new dense, fast-moving gas, which reforms thanks to radiative processes.
    \item These models also show that cooling-driven recondensation and hydrodynamic shielding promote the survival of dense gas and the development of multi-phase filamentary outflows. Re-condensation of mixed gas ensures that cold gas is replenished along the outflow, and hydrodynamic shielding reduces drag forces and contributes to preserving dense gas.
    \item Our wind-cloud models show that magnetic fields have significant effects on the morphology and spectral signatures of H\,{\sc i} gas. Magnetic fields, transverse to the wind, have two effects on clouds: they shield them by making the flow around them more laminar, and they squeeze them into a sheet-like configuration. Such morphological changes are evident in H\,{\sc i} column number densities.
    \item The morphological effects of transverse magnetic fields also leave clear imprints on synthetic spectral lines of H\,{\sc i} gas. Broader absorption spectra of H\,{\sc i} are characteristic of clouds embedded in transverse magnetic fields, compared to those in aligned fields.
\end{itemize}

Overall, our computer simulations can capture the formation and evolution of complex multi-phase outflows. Our simulations have shed light on how dense gas (e.g. H\,{\sc i}) can be preserved in these outflows and how magnetic fields can alter the properties of such dense gas and their resulting spectral signatures.

\textbf{Acknowledgements:} We thank the anonymous referee for their comments on our paper. The authors gratefully acknowledge the Gauss Centre for Supercomputing e.V. (\url{www.gauss-centre.eu}) for funding this project by providing computing time (via grant pn34qu) on the GCS Supercomputer SuperMUC-NG at the Leibniz Supercomputing Centre (\url{www.lrz.de}). In addition, the authors thank CEDIA (\url{www.cedia.edu.ec}) for providing access to their HPC cluster as well as for their technical support. We also thank the developers of the PLUTO code for making this hydrodynamic code available to the community.

\end{document}